\pdfoutput=1

\documentclass[%
 reprint,
 amsmath,amssymb,
 aps,
]{revtex4-2}

\usepackage{graphicx}
\usepackage{dcolumn}
\usepackage{bm}

\begin{document}

\preprint{APS/123-QED}

\title{Molecular Vibration Explorer: an online database and toolbox for surface-enhanced frequency conversion, infrared and Raman spectroscopy}

\author{Zsuzsanna Koczor-Benda}
\email{z.koczor-benda@ucl.ac.uk}
\affiliation{Department of Physics and Astronomy, University College London, London, WC1E 6BT, United Kingdom}
\affiliation{Department of Chemistry, King's College London, London, SE1 1DB, United Kingdom}

\author{Philippe Roelli}
\affiliation{Nano-optics group, CIC nanoGUNE BRTA, 20018 San Sebasti\'an, Spain}

\author{Christophe Galland}
\affiliation{Institute of Physics, Ecole Polytechnique F\'ed\'erale de Lausanne (EPFL), CH-1015 Lausanne, Switzerland}

\author{Edina Rosta} 
\affiliation{Department of Physics and Astronomy, University College London, London, WC1E 6BT, United Kingdom }
\affiliation{Department of Chemistry, King's College London, London, SE1 1DB, United Kingdom}

\date{\today}

\begin{abstract}
We present Molecular Vibration Explorer, a freely accessible online database and interactive tool for exploring vibrational spectra and tensorial light-vibration coupling strengths of a large collection of thiolated molecules. The `Gold' version of the database gathers the results from density functional theory calculations on 2'800 commercially available thiol compounds linked to a gold atom, with the main motivation to screen the best molecules for THz and mid-infrared to visible upconversion \cite{roelli_molecular_2020,koczor-benda_molecular_2021,chen_continuous-wave_2021,xomalis_detecting_2021}. Additionally, the `Thiol' version of the database contains results for 1'900 unbound thiolated compounds.
They both provide access to a comprehensive set of computed spectroscopic parameters for all vibrational modes of all molecules in the database. 
The user can simultaneously investigate infrared absorption, Raman scattering and vibrational sum- and difference frequency generation cross sections. 
Molecules can be screened for various parameters in custom frequency ranges, such as large Raman cross-section under specific molecular orientation, or large orientation-averaged sum-frequency generation (SFG) efficiency. 
The user can select polarization vectors for the electromagnetic fields, set the orientation of the molecule and customize parameters for plotting the corresponding IR, Raman and sum-frequency spectra. 
We illustrate the capabilities of this tool with selected applications in the field of surface-enhanced spectroscopy.

\end{abstract}

\maketitle


\section{Introduction}

Infrared (IR) absorption and Raman scattering spectroscopy constitute the most informative non-destructive optical methods to obtain fingerprints of molecular species.
While both signals are intrinsically too weak to allow detection of single molecules or molecular monolayers, surface-enhanced IR and Raman spectroscopy (SEIRA and SERS, respectively) circumvent this limitation by leveraging a combination of chemical and electromagnetic (plasmonic) enhancement factors.
The former describe the effect of orbital modifications and charge transfer on IR and Raman activities when a molecule binds to a metal (most often gold, or silver, copper, etc.).
The latter results from the enhanced local fields provided by localised surface plasmon resonances supported by metallic nanostructures. 
Over the years, the power of SEIRA and SERS for molecular spectroscopy has been firmly established, with applications extending beyond fundamental research into biosensing and security \cite{neubrech_surface-enhanced_2017,kneipp_serssingle-molecule_2008,fan_review_2020}.

IR absoprtion and Raman scattering can also be leveraged together in a process called vibrational sum-frequency generation \cite{chen_continuous-wave_2021}, which has proven useful for time-resolved studies of vibrational properties and dynamics of molecules at interfaces \cite{shen_fundamentals_2016,raschke_doubly-resonant_2002,roke_nonlinear_2012}, such as in the context of catalysis. 
Moreover, recent proposals and experiments inspired by the field of cavity optomechanics \cite{roelli_molecular_2016,roelli_molecular_2020} have translated this technique into a potential technology for frequency upconversion from the THz and IR domain into the visible domain \cite{chen_continuous-wave_2021,xomalis_detecting_2021}.
Another related and emerging research field is that of vibrational polaritons \cite{shalabney_coherent_2015,garcia-vidal_manipulating_2021,bylinkin_real-space_2021}, which aims in particular at remotely controlling chemical properties and reaction rates through strong coupling of vibrational modes with electromagnetic vacuum fluctuations. 

The two preceding paragraphs provide motivations for the work presented here, which consists in the development of an openly accessible and comprehensive numerical tool allowing to explore vibrational spectra and light-vibration coupling on thousands of commercially available molecules amenable to integration in gold plasmonic antennas and cavities. 
Indeed, many experimental groups lack either expertise, hardware, software, time or workforce -- or all of them -- for scouting for molecules with optimal properties for a specific goal (IR strong coupling, frequency upconversion, Raman tag, etc.).
Here, we present an open access, open source online platform called Molecular Vibration Explorer, which remedies this bottleneck and makes it easy to find molecular species featuring optimal properties for a specific application and specific geometry.
Among the main features available on our platform are:
\begin{itemize}
    \item For each molecule, orientation-averaged and orientation-dependent IR, Raman, and frequency conversion (SFG) spectra, as calculated by DFT. 
    \item Interactive sorting and screening of the entire database over user-defined frequency ranges according to a chosen parameter, such as IR or Raman cross-section. 
    \item 3D interactive plots of molecules and their vibrational modes, dynamically linked to the orientation-dependent values of IR and Raman cross-sections.
    \item Identification of chemically similar molecules within the database as well as from a list of known self-assembling molecules.
\end{itemize}

In the following, we first describe the theory and methodology behind the toolbox (Secs.~\ref{sec:theory} and \ref{sec:methods}), before presenting its main functionalities in Sec.~\ref{sec:functions}. We conclude by discussing examples of potential applications in Sec.~\ref{sec:appli}.

\section{Definitions}\label{sec:theory}

We calculate both orientation-averaged (denoted by angle brackets) and orientation-specific intensities for IR absorption, Raman scattering and frequency conversion (specifically, SFG) processes.
The IR absorption intensity (in [km mol$^{-1}$] units) of vibrational mode $m$ is defined as
\begin{equation}
 \langle I^\mathrm{A}_m \rangle=C^\mathrm{A} \langle |\underline{e}_\mathrm{IR} \hspace{0.1cm} \cdot \underline{\mu}'_m|^2  \rangle,
\label{eqIA}
\end{equation}
where $\underline{e}_\mathrm{IR}$ the field polarization vector of the IR beam, $\underline{\mu}'_m$ is the dipole derivative vector, and $C^\mathrm{A}=2.9254\times 10^{3}$ assuming $\underline{\mu}'_m$ is given in atomic units [e-bohr$^{2}$bohr$^{-2}$amu$^{-1}$].

The Raman intensity (differential Raman cross-section) for Stokes bands is given in [cm$^2$/sr] units 
\begin{align}
 \langle I^\mathrm{R}_m \rangle
= C^\mathrm{R} \frac{(\bar{\nu}_\mathrm{R} - \bar{\nu}_m)^4}{\bar{\nu}_m } \frac{1}{(1-\mathrm{exp}(-hc\bar{\nu}_m/k_BT))} \cdot \nonumber\\
\cdot \langle |\underline{e}_\mathrm{R,in} \hspace{0.1cm} \underline{ \underline{\alpha}}'_m \hspace{0.1cm} \underline{e}_\mathrm{R,out}|^2  \rangle, 
\label{eqIR}
\end{align}
where $\bar{\nu}_\mathrm{R}$ and  $\bar{\nu}_m$ are the wavenumbers of the pump laser and of the normal mode, respectively. The prefactor includes the effect of thermal occupancy of the vibration at temperature $T$. The polarization vectors of the pump (in) and Raman scattered (out) fields are given by $\underline{e}_\mathrm{R,in}$ and $\underline{e}_\mathrm{R,out}$, and $\underline{ \underline{\alpha}}'_m$ is the polarizability derivative tensor.
If $\underline{ \underline{\alpha}}'_m$ is given in [bohr$^4$amu$^{-1}$] units and wavenumbers in [cm$^{-1}$], the constant scaling factor is $C^\mathrm{R}=2.060 \times 10^{-45}$.

The SFG intensity is calculated as
\begin{equation}  
 \langle I^\mathrm{c}_m \rangle=C \frac{(\bar{\nu}_\mathrm{R} + \bar{\nu}_m)^4}{\bar{\nu}_m } \langle |\underline{e}_\mathrm{IR} \hspace{0.1cm} \cdot \underline{\mu}'_m|^2 |\underline{e}_\mathrm{R,in} \hspace{0.1cm} \underline{ \underline{\alpha}}'_m \hspace{0.1cm} \underline{e}_\mathrm{R,out}|^2 \rangle,
\label{eq:convI2}
\end{equation}
where $C=C^\mathrm{A}C^\mathrm{R}$. $I^\mathrm{c}_m $ measures the net increase in anti-Stokes Raman intensity due to IR pumping, hence the plus sign in the wavenumber dependent factor, and the omission of the Bose Einstein occupancy. 

Note that when averaging over multiple orientations, the overlap of the IR and Raman modes need to be considered for each orientation individually, as
\begin{align}
    \langle |\underline{e}_\mathrm{IR} \hspace{0.1cm} \cdot \underline{\mu}'_m|^2 & |\underline{e}_\mathrm{R,in} \hspace{0.1cm} \underline{ \underline{\alpha}}'_m \hspace{0.1cm} \underline{e}_\mathrm{R,out}|^2 \rangle \ne \nonumber\\ 
    &\langle |\underline{e}_\mathrm{IR} \hspace{0.1cm} \cdot \underline{\mu}'_m|^2 \rangle \langle |\underline{e}_\mathrm{R,in} \hspace{0.1cm} \underline{ \underline{\alpha}}'_m \hspace{0.1cm} \underline{e}_\mathrm{R,out}|^2 \rangle.
\end{align}
Nevertheless, analytic formulas for calculating the averages over all possible relative orientations of the molecule and fields, considering all possible field polarization settings, can be derived (cf. Supplementary Material).
For orientation-specific intensities the angle brackets in Eqs. \ref{eqIA} - \ref{eq:convI2} can be omitted.

Apart from individual normal mode intensities, the target properties $A$, $R$, and $P$ defined below  can also be used for ranking the molecules of the database in a specific frequency range. 
\begin{equation}
A=\mathrm{log} \Big( \sum_{m \in M} 
 \langle I^\mathrm{A}_m \rangle \Big),  
\label{eq:targetA}
\end{equation}

\begin{equation}
R=\mathrm{log} \Big( \sum_{m \in M} 
 \langle I^\mathrm{R}_m \rangle \Big),  
\label{eq:targetR}
\end{equation}

\begin{equation}
P=\mathrm{log} \Big( \sum_{m \in M} 
 \langle I^\mathrm{c}_m \rangle \Big),  
\label{eq:targetP}
\end{equation}
where $M$ refers to the set of normal modes that have frequencies in the range defined by the user. These properties are standardized as described in Ref \cite{koczor-benda_molecular_2021}. $A$, $R$, and $P$ can also be calculated using orientation-specific intensities, this enables fast comparison of molecular performance at different orientations.  

The vibrational spectra of each molecule can be plotted as discrete (stick) spectra or broadened spectra. In the latter case, a Lorentzian broadening is applied to IR and Raman spectra, for which FWHM$_\mathrm{IR}$ and FWHM$_\mathrm{R}$ can be specified separately, and the SFG spectrum inherits the lineshape of the product of these two Lorentzians.

\section{Methods}\label{sec:methods}
The creation of the database is described in Ref. \cite{koczor-benda_molecular_2021}. DFT calculations are performed at the B3LYP+D3/def2-SVP level using the Gaussian program package. Analytic formulas for the orientation averages of intensities were derived using Mathematica.
The Molecular Vibration Explorer can be accessed via the Materials Cloud platform.
The web application is created using Voila, and is based on interactive Jupyter notebooks.
Molecular similarity scores are generated using RDKit. NGLview is used for the 3D visualization of molecules.

\section{Functionalities}\label{sec:functions}

\subsection{Exploring the databases}

The \textit{Gold} and \textit{Thiol} databases can be explored separately.
The database page allows for ranking molecules within the database and selecting them for further analysis. 
The user may choose the \textit{target property} ($A$,$R$, or $P$) and set a frequency range of interest. 
A global frequency scaling factor can be applied to improve agreement between DFT calculated and experimentally measured frequencies.
The respective directions of the field polarization vectors ($\underline{e}_\mathrm{IR}$, $\underline{e}_\mathrm{R,in}$, and $\underline{e}_\mathrm{R,out}$) can be specified independently, among three possible orthogonal directions $x,y,z$. Intensities $\langle I^\mathrm{A}_m \rangle$, $\langle I^\mathrm{R}_m \rangle$, and $\langle I^\mathrm{c}_m \rangle$ have been pre-computed and stored for all possible combinations of field polarizations and considering a random orientation of molecules. 
A temperature of 298.15~K and Raman laser wavelength of 785~nm were used for computing Raman and conversion intensities. 
The distribution of the target property across the entire database, for mode frequencies within the user-specified range, is plotted as a histogram, and a table is generated below that shows all molecules sorted according to decreasing value of the target. 
Further properties of the molecules and their individual normal modes can be explored by navigating the links displayed in the table, as described below.


\subsection{Molecular properties}

Upon clicking on ``\textit{Go to molecule page}'' next to a particular molecule, the new page displays vibrational spectra and other relevant molecular properties and enables a high level of customization.
In the tab labeled ``\textit{Set molecule orientation}'' on the left, the molecular orientation can be modified at will by specifying the rotational angles $\phi$, $\theta$, and $\xi$, with respect to the Cartesian coordinate system defined along the possible polarisation vectors.
The resulting orientation is instantly displayed on the 3D molecular viewer while the corresponding Raman, IR and SFG spectra are also updated. 
The spectra for the selected orientation can be directly compared with those for the orientation average by checking ``\textit{Show full orientation average}'' in the ``\textit{Set plotting parameters}'' tab on the left. 
For Raman scattering and SFG the temperature and laser wavelength can be set in the  ``\textit{Set experimental parameters}'' tab, while the frequency range, scaling factor, style of the spectrum (stick, broadened, stick+broadened) and corresponding broadening parameters (FWHM$_\mathrm{IR}$, FWHM$_\mathrm{R}$) can all be tuned in the  ``\textit{Set plotting parameters}'' tab. 
Finally, field polarization directions can be specified in the ``\textit{Set field polarization}'' tab. 

In addition to optical properties, physical properties of the molecule relevant to surface-enhanced spectroscopy are listed below the 3D drawing. 
The length of the projection of the molecule along the $z$ axis is referred to as ``\textit{Layer height}''. Indeed, it corresponds to the approximate thickness of a molecular monolayer assuming binding through the thiol group onto a flat gold surface spanning the $x,y$ plane. 
Its geometrical projection on the flat gold surface is given by ``\textit{XY projection}''. 
Another feature that is especially useful for navigating the database is the table of similar molecules that is displayed at the bottom of the page. Molecules susceptible to form self-assembled monolayers  (SAMs) and similar to the chosen molecule are listed to help optimisation of surface-enhanced applications. 


\subsection{Properties of vibrational modes}

Upon clicking on ``\textit{Check normal modes}'' next to a particular molecule, the user can explore all normal modes of the selected molecule through 3D visualization. The dependence of the IR, Raman, and conversion/SFG intensity of a specific normal mode on molecular orientation is represented by a projection onto the plane perpendicular to the axis of rotation. 
The direction of the field polarization vectors of the IR and Raman (excitation and scattered) beams can also be set on this page.


\section{Example Applications}\label{sec:appli}
\begin{figure}[h!]
    \vspace{-12pt}
    \centering
    \includegraphics[width=1.0\columnwidth]{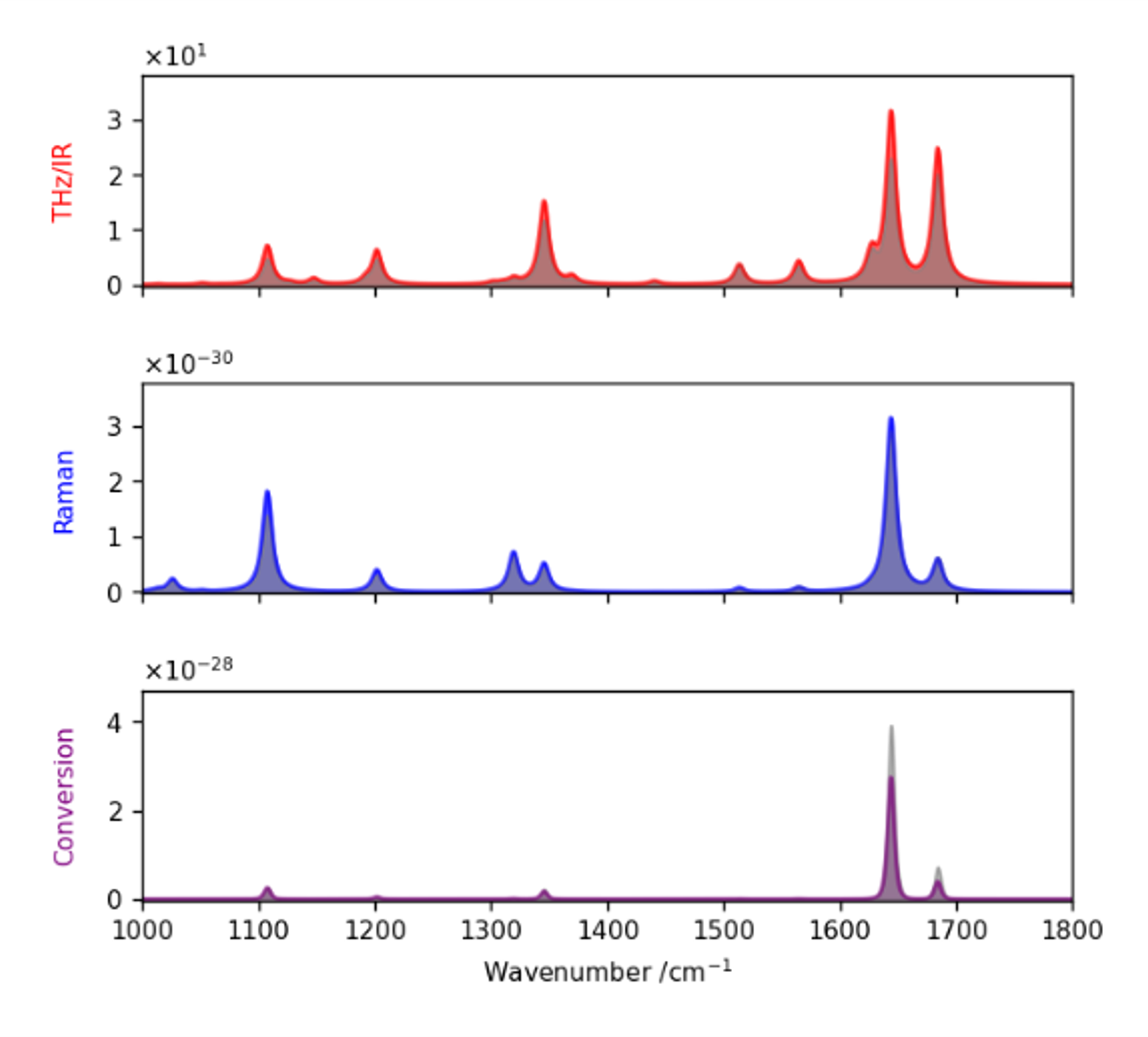}
    \vspace{-24pt}
    \caption{
      \begin{footnotesize}
    From top to bottom, numerical simulations of IR absorption (red line), Raman scattering (blue) and conversion (SFG, purple) spectra of a NH$_2$-BPhT molecule for a specific orientation. The orientation averaged spectra is depicted with grey areas for each spectrum. All local fields are chosen to be parallel to the z-axis. 
      \end{footnotesize}
      }
    \label{fig:spectra}
\end{figure}

Sum-Frequency generation (SFG) has proven to be a useful 
tool for a variety of investigations in surface science \cite{shen_fundamentals_2016}. 
The combined selection rules of IR and Raman processes enable 
to optically study the symmetry of crystalline structures \cite{liu_sum-frequency_2008}, 
to investigate the bonding of molecules to substrates \cite{cremer_ethylene_1996}
and the orientation of molecular layers on different surfaces \cite{hunt_observation_1987,roke_vibrational_2003}. 
The experimental characterization of these molecules requires access to the non-vanishing elements of the non-linear polarizability tensor, which can for certain molecules be addressed sequentially using different polarization combinations. 
All these components can be extracted with the help of our toolbox and can be used to reconstruct the IR, Raman and upconversion (SFG) spectra for all combinations of local electromagnetic fields and orientation of the molecule.
Figure~\ref{fig:spectra} illustrates different spectra obtained for the NH$_2$-BPhT molecule both for a specific orientation of the molecule and for the orientation-averaged case. In this section of the manuscript, the specific orientation will correspond to the initial orientation of the Gold database's molecules in the Molecular Vibration Explorer (MVE toolbox): SH (thiol) bond along the $z$-axis, while the polarization of the three local electromagnetic fields (IR, Raman excitation, Raman scattering), if not explicitly given in the text, will be chosen along $z$. 
This default configuration is expected to well approximate most common situations in surface-enhanced studies.
One interesting feature for spectroscopy is that the more constrained selection rules for SFG allow only a few vibrational modes to feature sizeable upconversion efficiency. This aspect can be leveraged in experiments where the level of background arising from the substrates or other molecules present in the vicinity might prevent detection of the targeted molecules.



\begin{figure}[h!]
    \vspace{-12pt}
    \centering
    \includegraphics[width=1.0\columnwidth]{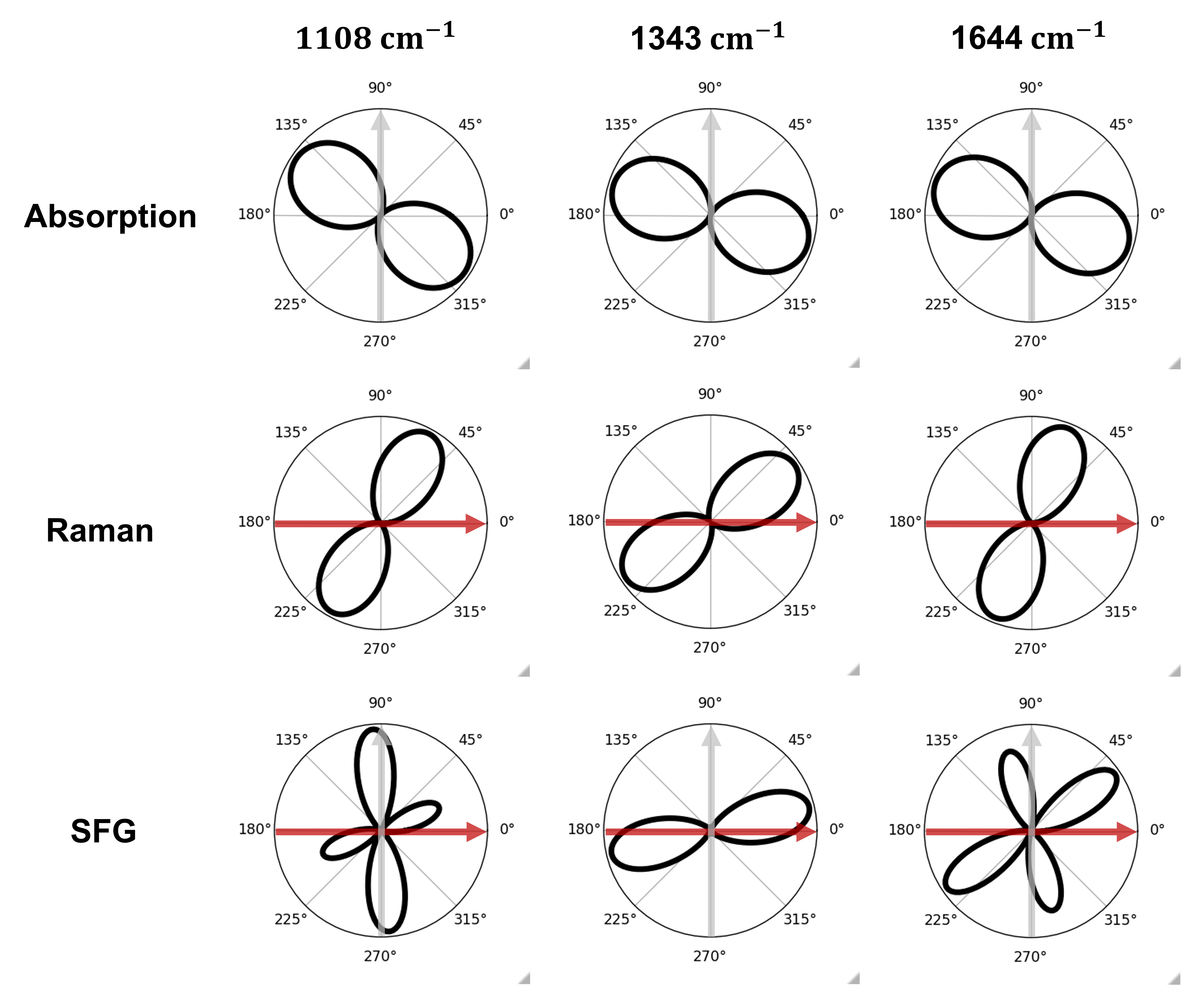}
    \vspace{-18pt}
    \caption{
      \begin{footnotesize}
    From top to bottom, numerical simulations of the dependence of the IR absorption, Raman scattering and SFG spectra on the orientation of the molecule with respect to external local fields for the three main upconversion-active modes of NH$_2$-BPhT. In this figure, local fields are chosen to be cross-polarized: IR field along y (gray arrow), Raman in/out fields along x (red arrow).  
      \end{footnotesize}
      }
    \label{fig:orientation}
\end{figure}

In the situation where the polarizations are defined by a plasmonic nanostructure and a specific orientation of the molecule is expected \cite{chen_continuous-wave_2021},  the MVE toolbox permits to investigate the orientation dependence of the IR absorption,  Raman and SFG signals. 
Figure~\ref{fig:orientation} illustrates, for the case of cross-polarized local fields (IR field along y, Raman in/out fields along x), the changes in the different processes' magnitudes as function of the molecule orientation. 
As evidenced in the figure, even though the corresponding absorption and Raman signatures look rather similar, the molecular orientation dependence of SFG efficiency shows clearly distinct patterns for different vibrational modes of NH$_2$-BPhT.
We note that a cross-polarization configuration is not typical in doubly resonant plasmonic cavities but could be achieved using specific antennas \cite{schnell_phase-resolved_2010}. 
Thanks to the prediction of orientation dependence the MVE toolbox may help to optically assess the orientation of self-assembled monolayers within metallic structures \cite{ahmed_structural_2021} in a similar way as it could be determined in TERS experiments \cite{zhang_chemical_2013}. 
The MVE toolbox will also be useful in designing nanostructures and molecular layers maximizing the conversion efficiency of the SFG process for frequency conversion applications \cite{roelli_molecular_2020,chen_continuous-wave_2021}. 

\subsection{Molecule optimisation or fingerprinting} 
\begin{figure}
    \centering
    \vspace{-10pt}
    \includegraphics[width=0.95\columnwidth]{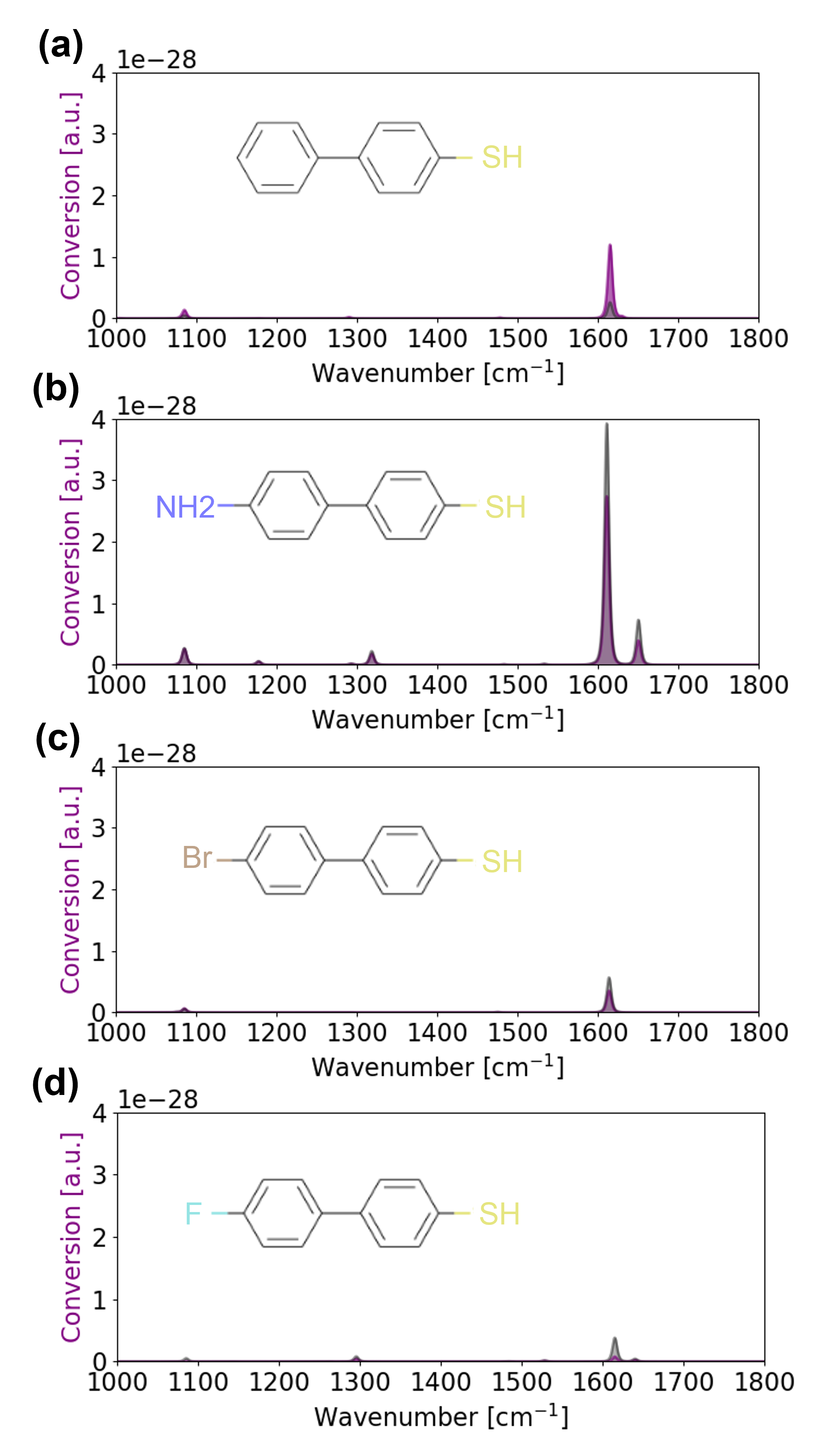}
    \vspace{-10pt}
    \caption{
      \begin{footnotesize}
    Averaged (grey area) and orientation specific (purple line) conversion spectra of BPhT \textbf{(a)}, NH$_2$-BPhT \textbf{(b)}, 
    Br-BPhT \textbf{(c)} and F-BPhT \textbf{(d)}. 
      \end{footnotesize}
      }
    \label{fig:link}
\end{figure}

Another opportunity offered by the MVE database is to scrutinize molecules similar to a reference molecule already controlled experimentally. As demonstrated in Fig.~\ref{fig:link}, a simple change in one termination of the BPhT molecule can lead to significant changes in the respective magnitude of the different SFG active modes. 
The toolbox therefore provides a way to optimize upconversion efficiency by carefully designing the molecule and could also be used reversely in order to experimentally differentiate the upconverted signals from very similar molecules, changing only by one termination \cite{zhang_vibrational_1994}. 

\subsection{SERS and advanced IR detection} 
\begin{figure}[h!]
    \centering
    \vspace{-12pt}
    \includegraphics[width=1.0\columnwidth]{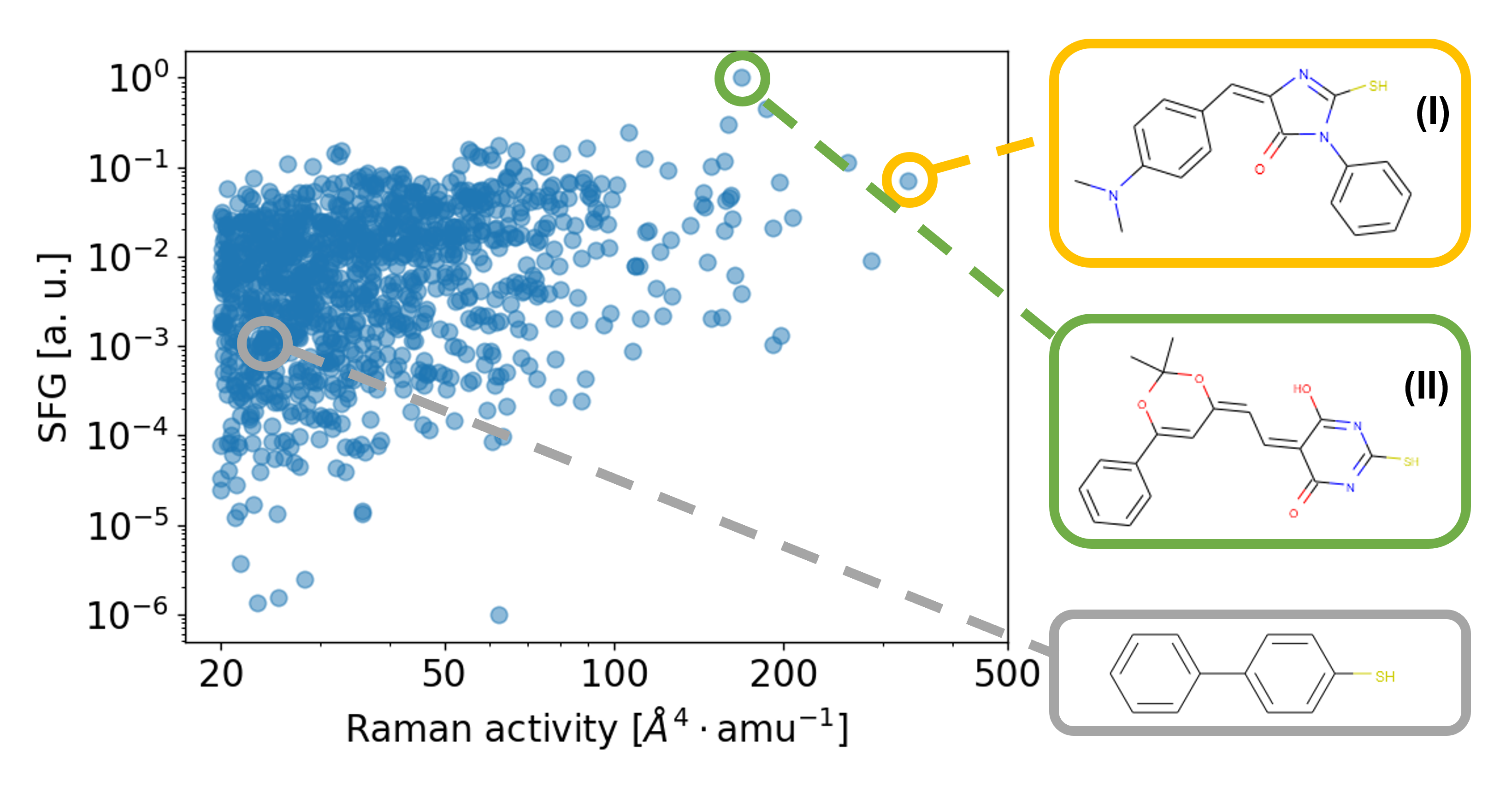}
    \vspace{-18pt}
    \caption{
      \begin{footnotesize}
    Distribution of SFG coefficients and Raman activities for all vibrational modes of the Gold database lying in the $20-50$~THz spectral range, identified in Ref.~\cite{roelli_molecular_2020} for its interest in IR detector applications. We highlight BPhT with a grey circle, the only molecule ever used to date in a continuous wave SFG experiment~\cite{chen_continuous-wave_2021}. Two other molecules with significantly higher calculated SFG efficiencies are highlighted in yellow and green. Full names of the molecules are given in the SM.
      \end{footnotesize}
      }
    \label{fig:sers}
\end{figure} 

One direct application of the MVE database consists in exploring possible molecules for improved SERS signals in applications such as SERS tags \cite{wang_sers_2013} or fundamental studies of molecular optomechanics \cite{roelli_molecular_2016}. 
As evidenced in Fig.~\ref{fig:sers}, there are numerous commercially available molecules compatible with gold functionalization that would reach levels of Raman activity comparable or higher than standard compounds used in SERS experiments like BPhT. 
Additionally, filtering the database on the region of interest for low-light IR detectors at room temperature \cite{roelli_molecular_2020} (20-50 THz), we can identify molecular vibrations with SFG coefficients orders of magnitude larger than the one of the molecule (BPhT) used in the recent surface-enhanced upconversion experiments \cite{chen_continuous-wave_2021, xomalis_detecting_2021}. 

\subsection{IR strong coupling}
\begin{figure}[h!]
    \centering
    \vspace{-24pt}
    \includegraphics[width=1.\columnwidth]{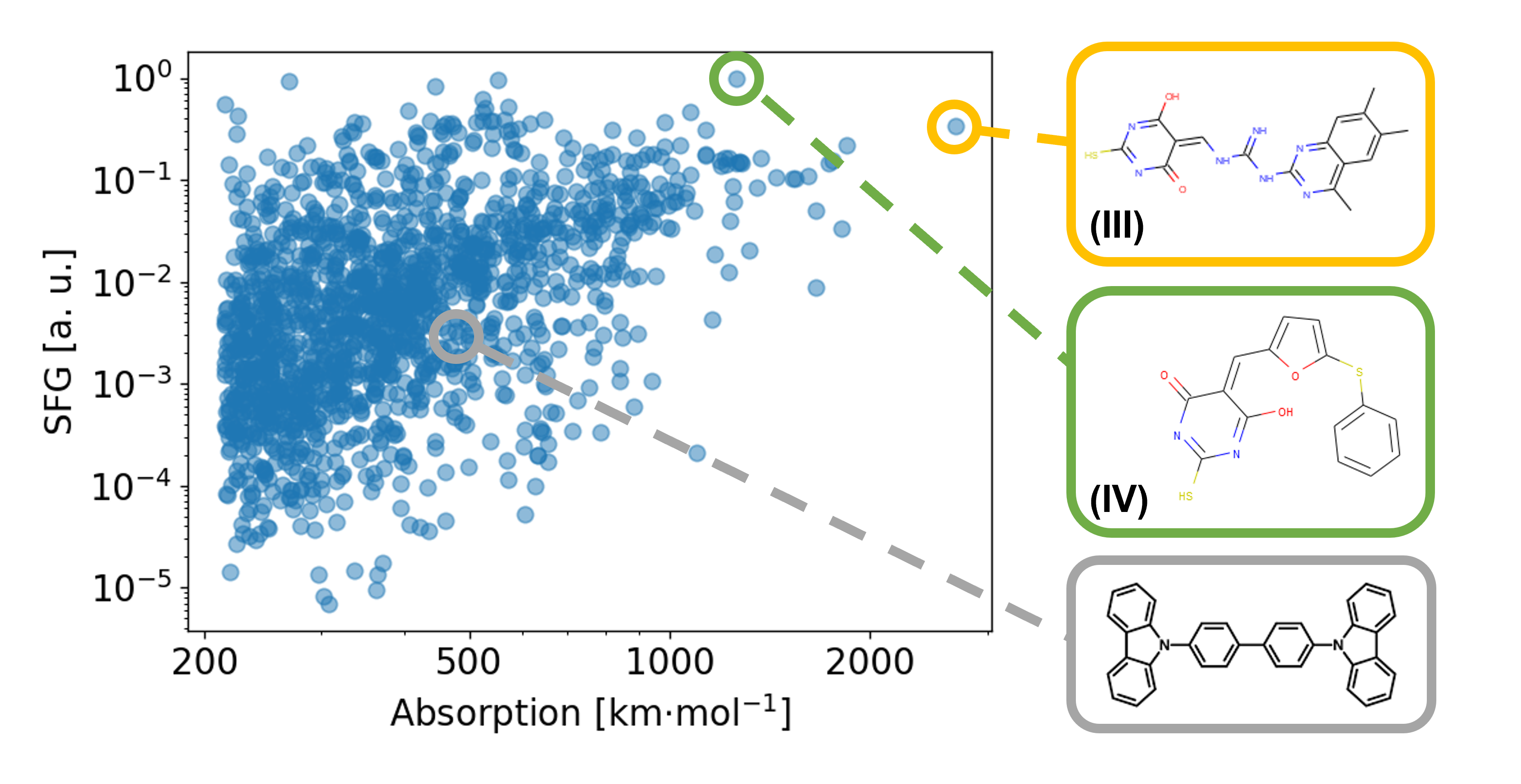}
    \vspace{-18pt}
    \caption{
      \begin{footnotesize}
    Distribution of SFG coefficients and Raman activities for all vibrational modes of the Gold database lying in the $1400-1550$~cm$^{-1}$ range that match surface phonon-polariton (SPhP) modes of hexagonal boron nitride (hBN). 
    We highlight with a grey circle the molecule (CBP)  experimentally used to achieve vibrational strong coupling with an SPhP mode of hBN~\cite{bylinkin_real-space_2021}.  Two other molecules with significantly higher calculated IR absorption cross sections and SFG efficiencies are highlighted in yellow and green. Full names of the molecules are given in the SM.
      \end{footnotesize}
      }
    \label{fig:sc}
\end{figure}

The MVE database also opens new routes to explore the regime of IR-vibration strong coupling \cite{garcia-vidal_manipulating_2021} in systems containing one or few molecules only. 
Recent studies of the coupling between molecular vibrations and surface phonon-polaritons modes 
of hBN have achieved this regime \cite{bylinkin_real-space_2021}. 
The narrow range accessible with these phonon-polaritons modes [1400-1550 cm$^{-1}$] requires very specific molecules 
to achieve sufficiently strong coupling. The Molecular Vibration Explorer 
enables to quickly identify molecules (Fig.~\ref{fig:sc}) with IR cross sections that are promising for strong coupling applications. 
In addition, the database allows to identify vibrational modes that would not only reach the IR-vibration strong-coupling regime but would simultaneously enable substantial conversion of the generated phonons to the visible via molecular optomechanical upconversion. 
The molecule used in Ref. \cite{bylinkin_real-space_2021} is illustrated on the same figure to highlight the number of other molecular vibrations that would achieve similar or better IR intensities while enabling more suitable or controlled deposition, smaller footprint or higher compatibility with molecular SFG experiments. 
The same approach could also be instrumental in the design of the interaction between other surface modes like graphene plasmons and vibrations 
\cite{rodrigo_mid-infrared_2015,epstein_far-field_2020}. 

\subsection{Molecule detection} 
\begin{figure*}[t!]
    \centering
    \vspace{-12pt}
    \includegraphics[width=.85\textwidth]{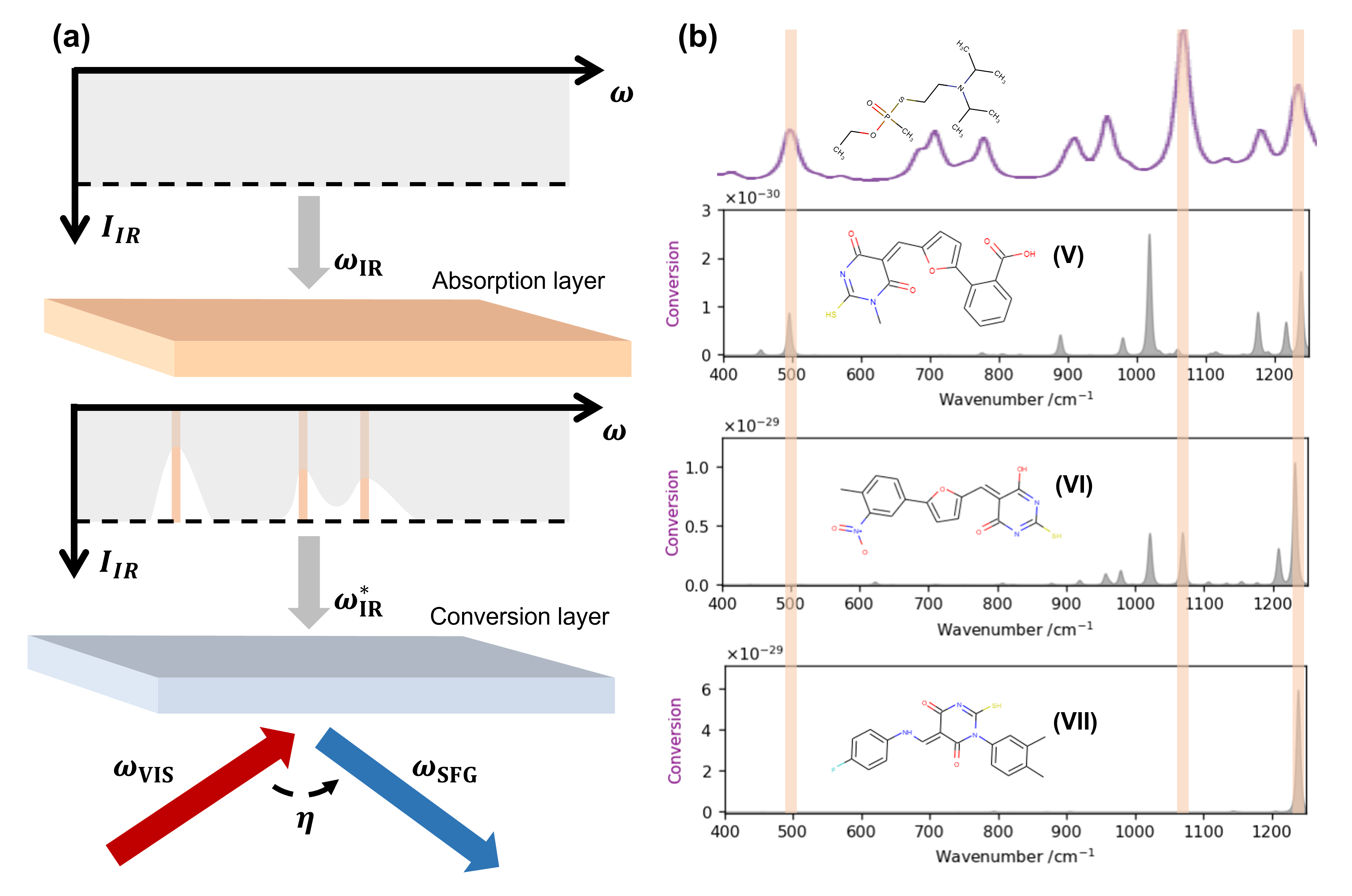}
    \vspace{-12pt}
    \caption{
      \begin{footnotesize}
        \textbf{(a)} Schematics of the proposed detection scheme for substance identification. From top to bottom, broadband infrared light is passing through the substance to be identified \cite{tittl_imaging-based_2018}, after which it is analysed thanks to an upconversion layer containing molecules with SFG active modes at matched frequencies, so that spectroscopy is eventually performed using visible light detectors. 
        \textbf{(b)} Example of SFG active molecules identified in the database with vibrational frequencies matching the main characteristic IR absorption lines of the VX nerve agent. The top spectrum corresponds to the numerically calculated absorption spectrum of the VX nerve agent (adapted from Ref.~\cite{mott_calculated_2012}). Then, from top to bottom, the conversion spectra correspond to the spectra extracted from the MVE database for the compounds V, VI and VII. Full name of the molecules are given in the SM.
      \end{footnotesize}
      }
    \label{fig:det}
\end{figure*}

One prospective technological application of the database is related to the sub-wavelength dimension of the newly demonstrated SFG devices \cite{chen_continuous-wave_2021}. 
In advanced designs, constituted of different sub-wavelength frequency upconverters, simultaneous detection at different frequencies would be possible, akin to metapixel imaging systems realised with dielectric nanostructures \cite{tittl_imaging-based_2018}. 
Such IR recognition devices could be valuable for the detection of toxic substances. 
For example, the VX nerve agent has different characteristic IR absorption lines \cite{mott_calculated_2012} that could well be targeted by SFG nanodevices. 
Figure~\ref{fig:det} shows different molecules extracted from the database that could efficiently detect variations of IR absorption at the {490, 1067, 1232 cm$^{-1}$} characteristic lines of the VX nerve agent. 

\subsection{Conclusion}

In conclusion, we presented the Molecular Vibrational Explorer (hosted on the Materials Cloud open platform) that allows interactive access to a large database (thousands of molecules) of vibrational and spectroscopic properties. 
The database targets surface-enhanced applications such as SERS, SEIRA and vibration-assisted frequency upconversion; it can also be expanded in the future with more molecules. 
We illustrated the use of the Molecular Vibrational Explorer in the optimisation of molecules for a number of applications including SERS tags, vibrational strong coupling, and frequency upconversion for substance detection. 
We believe that the Molecular Vibrational Explorer will foster progress and productivity in a broad range of research fields related to molecular spectroscopy, nano-optics and surface science.


\begin{acknowledgments}
ZKB thanks Balint Koczor for technical help with the analytic integration of orientation averages.
We acknowledge funding from the European Research Council (ERC) under Horizon 2020 research and innovation programme THOR (Grant Agreement No. 829067) and QTONE (Grant Agreement No. 820196). 
ZKB and ER also acknowledge funding from EPSRC (EP/R013012/1, EP/L027151/1) and ERC project 757850 BioNet. We are grateful to the UK Materials and Molecular Modelling Hub for computational resources, which is partially funded by EPSRC (EP/P020194/1). PR acknowledges financial support from the Spanish Ministry of Science, Innovation and Universities (national project RTI2018-094830-B-100 and project MDM-2016-0618 of the Maria de Maeztu Units of Excellence Program). C.G. acknowledges support from the Swiss National Science Foundation (project nos. 170684 and 198898).
\end{acknowledgments}


\bibliography{paper}

\begin{thebibliography}{27}%
\makeatletter
\providecommand \@ifxundefined [1]{%
 \@ifx{#1\undefined}
}%
\providecommand \@ifnum [1]{%
 \ifnum #1\expandafter \@firstoftwo
 \else \expandafter \@secondoftwo
 \fi
}%
\providecommand \@ifx [1]{%
 \ifx #1\expandafter \@firstoftwo
 \else \expandafter \@secondoftwo
 \fi
}%
\providecommand \natexlab [1]{#1}%
\providecommand \enquote  [1]{``#1''}%
\providecommand \bibnamefont  [1]{#1}%
\providecommand \bibfnamefont [1]{#1}%
\providecommand \citenamefont [1]{#1}%
\providecommand \href@noop [0]{\@secondoftwo}%
\providecommand \href [0]{\begingroup \@sanitize@url \@href}%
\providecommand \@href[1]{\@@startlink{#1}\@@href}%
\providecommand \@@href[1]{\endgroup#1\@@endlink}%
\providecommand \@sanitize@url [0]{\catcode `\\12\catcode `\$12\catcode
  `\&12\catcode `\#12\catcode `\^12\catcode `\_12\catcode `\%12\relax}%
\providecommand \@@startlink[1]{}%
\providecommand \@@endlink[0]{}%
\providecommand \url  [0]{\begingroup\@sanitize@url \@url }%
\providecommand \@url [1]{\endgroup\@href {#1}{\urlprefix }}%
\providecommand \urlprefix  [0]{URL }%
\providecommand \Eprint [0]{\href }%
\providecommand \doibase [0]{https://doi.org/}%
\providecommand \selectlanguage [0]{\@gobble}%
\providecommand \bibinfo  [0]{\@secondoftwo}%
\providecommand \bibfield  [0]{\@secondoftwo}%
\providecommand \translation [1]{[#1]}%
\providecommand \BibitemOpen [0]{}%
\providecommand \bibitemStop [0]{}%
\providecommand \bibitemNoStop [0]{.\EOS\space}%
\providecommand \EOS [0]{\spacefactor3000\relax}%
\providecommand \BibitemShut  [1]{\csname bibitem#1\endcsname}%
\let\auto@bib@innerbib\@empty
\bibitem [{\citenamefont {Roelli}\ \emph {et~al.}(2020)\citenamefont {Roelli},
  \citenamefont {Martin-Cano}, \citenamefont {Kippenberg},\ and\ \citenamefont
  {Galland}}]{roelli_molecular_2020}%
  \BibitemOpen
  \bibfield  {author} {\bibinfo {author} {\bibfnamefont {P.}~\bibnamefont
  {Roelli}}, \bibinfo {author} {\bibfnamefont {D.}~\bibnamefont {Martin-Cano}},
  \bibinfo {author} {\bibfnamefont {T.~J.}\ \bibnamefont {Kippenberg}},\ and\
  \bibinfo {author} {\bibfnamefont {C.}~\bibnamefont {Galland}},\ }\bibfield
  {title} {\bibinfo {title} {Molecular {Platform} for {Frequency}
  {Upconversion} at the {Single}-{Photon} {Level}},\ }\href
  {https://doi.org/10.1103/PhysRevX.10.031057} {\bibfield  {journal} {\bibinfo
  {journal} {Physical Review X}\ }\textbf {\bibinfo {volume} {10}},\ \bibinfo
  {pages} {031057} (\bibinfo {year} {2020})}\BibitemShut {NoStop}%
\bibitem [{\citenamefont {Koczor-Benda}\ \emph {et~al.}(2021)\citenamefont
  {Koczor-Benda}, \citenamefont {Boehmke}, \citenamefont {Xomalis},
  \citenamefont {Arul}, \citenamefont {Readman}, \citenamefont {Baumberg},\
  and\ \citenamefont {Rosta}}]{koczor-benda_molecular_2021}%
  \BibitemOpen
  \bibfield  {author} {\bibinfo {author} {\bibfnamefont {Z.}~\bibnamefont
  {Koczor-Benda}}, \bibinfo {author} {\bibfnamefont {A.~L.}\ \bibnamefont
  {Boehmke}}, \bibinfo {author} {\bibfnamefont {A.}~\bibnamefont {Xomalis}},
  \bibinfo {author} {\bibfnamefont {R.}~\bibnamefont {Arul}}, \bibinfo {author}
  {\bibfnamefont {C.}~\bibnamefont {Readman}}, \bibinfo {author} {\bibfnamefont
  {J.~J.}\ \bibnamefont {Baumberg}},\ and\ \bibinfo {author} {\bibfnamefont
  {E.}~\bibnamefont {Rosta}},\ }\bibfield  {title} {\bibinfo {title} {Molecular
  {Screening} for {Terahertz} {Detection} with {Machine}-{Learning}-{Based}
  {Methods}},\ }\href {https://doi.org/10.1103/PhysRevX.11.041035} {\bibfield
  {journal} {\bibinfo  {journal} {Physical Review X}\ }\textbf {\bibinfo
  {volume} {11}},\ \bibinfo {pages} {041035} (\bibinfo {year}
  {2021})}\BibitemShut {NoStop}%
\bibitem [{\citenamefont {Chen}\ \emph {et~al.}(2021)\citenamefont {Chen},
  \citenamefont {Roelli}, \citenamefont {Hu}, \citenamefont {Verlekar},
  \citenamefont {Amirtharaj}, \citenamefont {Barreda}, \citenamefont
  {Kippenberg}, \citenamefont {Kovylina}, \citenamefont {Verhagen},
  \citenamefont {Martínez},\ and\ \citenamefont
  {Galland}}]{chen_continuous-wave_2021}%
  \BibitemOpen
  \bibfield  {author} {\bibinfo {author} {\bibfnamefont {W.}~\bibnamefont
  {Chen}}, \bibinfo {author} {\bibfnamefont {P.}~\bibnamefont {Roelli}},
  \bibinfo {author} {\bibfnamefont {H.}~\bibnamefont {Hu}}, \bibinfo {author}
  {\bibfnamefont {S.}~\bibnamefont {Verlekar}}, \bibinfo {author}
  {\bibfnamefont {S.~P.}\ \bibnamefont {Amirtharaj}}, \bibinfo {author}
  {\bibfnamefont {A.~I.}\ \bibnamefont {Barreda}}, \bibinfo {author}
  {\bibfnamefont {T.~J.}\ \bibnamefont {Kippenberg}}, \bibinfo {author}
  {\bibfnamefont {M.}~\bibnamefont {Kovylina}}, \bibinfo {author}
  {\bibfnamefont {E.}~\bibnamefont {Verhagen}}, \bibinfo {author}
  {\bibfnamefont {A.}~\bibnamefont {Martínez}},\ and\ \bibinfo {author}
  {\bibfnamefont {C.}~\bibnamefont {Galland}},\ }\bibfield  {title} {\bibinfo
  {title} {Continuous-wave frequency upconversion with a molecular
  optomechanical nanocavity},\ }\href {https://doi.org/10.1126/science.abk3106}
  {\bibfield  {journal} {\bibinfo  {journal} {Science}\ }\textbf {\bibinfo
  {volume} {374}},\ \bibinfo {pages} {1264} (\bibinfo {year}
  {2021})}\BibitemShut {NoStop}%
\bibitem [{\citenamefont {Xomalis}\ \emph {et~al.}(2021)\citenamefont
  {Xomalis}, \citenamefont {Zheng}, \citenamefont {Chikkaraddy}, \citenamefont
  {Koczor-Benda}, \citenamefont {Miele}, \citenamefont {Rosta}, \citenamefont
  {Vandenbosch}, \citenamefont {Martínez},\ and\ \citenamefont
  {Baumberg}}]{xomalis_detecting_2021}%
  \BibitemOpen
  \bibfield  {author} {\bibinfo {author} {\bibfnamefont {A.}~\bibnamefont
  {Xomalis}}, \bibinfo {author} {\bibfnamefont {X.}~\bibnamefont {Zheng}},
  \bibinfo {author} {\bibfnamefont {R.}~\bibnamefont {Chikkaraddy}}, \bibinfo
  {author} {\bibfnamefont {Z.}~\bibnamefont {Koczor-Benda}}, \bibinfo {author}
  {\bibfnamefont {E.}~\bibnamefont {Miele}}, \bibinfo {author} {\bibfnamefont
  {E.}~\bibnamefont {Rosta}}, \bibinfo {author} {\bibfnamefont {G.~A.~E.}\
  \bibnamefont {Vandenbosch}}, \bibinfo {author} {\bibfnamefont
  {A.}~\bibnamefont {Martínez}},\ and\ \bibinfo {author} {\bibfnamefont
  {J.~J.}\ \bibnamefont {Baumberg}},\ }\bibfield  {title} {\bibinfo {title}
  {Detecting mid-infrared light by molecular frequency upconversion in
  dual-wavelength nanoantennas},\ }\href
  {https://doi.org/10.1126/science.abk2593} {\bibfield  {journal} {\bibinfo
  {journal} {Science}\ }\textbf {\bibinfo {volume} {374}},\ \bibinfo {pages}
  {1268} (\bibinfo {year} {2021})},\ \bibinfo {note} {publisher: American
  Association for the Advancement of Science}\BibitemShut {NoStop}%
\bibitem [{\citenamefont {Neubrech}\ \emph {et~al.}(2017)\citenamefont
  {Neubrech}, \citenamefont {Huck}, \citenamefont {Weber}, \citenamefont
  {Pucci},\ and\ \citenamefont {Giessen}}]{neubrech_surface-enhanced_2017}%
  \BibitemOpen
  \bibfield  {author} {\bibinfo {author} {\bibfnamefont {F.}~\bibnamefont
  {Neubrech}}, \bibinfo {author} {\bibfnamefont {C.}~\bibnamefont {Huck}},
  \bibinfo {author} {\bibfnamefont {K.}~\bibnamefont {Weber}}, \bibinfo
  {author} {\bibfnamefont {A.}~\bibnamefont {Pucci}},\ and\ \bibinfo {author}
  {\bibfnamefont {H.}~\bibnamefont {Giessen}},\ }\bibfield  {title} {\bibinfo
  {title} {Surface-{Enhanced} {Infrared} {Spectroscopy} {Using} {Resonant}
  {Nanoantennas}},\ }\href {https://doi.org/10.1021/acs.chemrev.6b00743}
  {\bibfield  {journal} {\bibinfo  {journal} {Chemical Reviews}\ }\textbf
  {\bibinfo {volume} {117}},\ \bibinfo {pages} {5110} (\bibinfo {year}
  {2017})}\BibitemShut {NoStop}%
\bibitem [{\citenamefont {Kneipp}\ \emph {et~al.}(2008)\citenamefont {Kneipp},
  \citenamefont {Kneipp},\ and\ \citenamefont
  {Kneipp}}]{kneipp_serssingle-molecule_2008}%
  \BibitemOpen
  \bibfield  {author} {\bibinfo {author} {\bibfnamefont {J.}~\bibnamefont
  {Kneipp}}, \bibinfo {author} {\bibfnamefont {H.}~\bibnamefont {Kneipp}},\
  and\ \bibinfo {author} {\bibfnamefont {K.}~\bibnamefont {Kneipp}},\
  }\bibfield  {title} {\bibinfo {title} {{SERS}—a single-molecule and
  nanoscale tool for bioanalytics},\ }\href {https://doi.org/10.1039/B708459P}
  {\bibfield  {journal} {\bibinfo  {journal} {Chemical Society Reviews}\
  }\textbf {\bibinfo {volume} {37}},\ \bibinfo {pages} {1052} (\bibinfo {year}
  {2008})}\BibitemShut {NoStop}%
\bibitem [{\citenamefont {Fan}\ \emph {et~al.}(2020)\citenamefont {Fan},
  \citenamefont {Andrade},\ and\ \citenamefont {Brolo}}]{fan_review_2020}%
  \BibitemOpen
  \bibfield  {author} {\bibinfo {author} {\bibfnamefont {M.}~\bibnamefont
  {Fan}}, \bibinfo {author} {\bibfnamefont {G.~F.~S.}\ \bibnamefont
  {Andrade}},\ and\ \bibinfo {author} {\bibfnamefont {A.~G.}\ \bibnamefont
  {Brolo}},\ }\bibfield  {title} {\bibinfo {title} {A review on recent advances
  in the applications of surface-enhanced {Raman} scattering in analytical
  chemistry},\ }\href {https://doi.org/10.1016/j.aca.2019.11.049} {\bibfield
  {journal} {\bibinfo  {journal} {Analytica Chimica Acta}\ }\textbf {\bibinfo
  {volume} {1097}},\ \bibinfo {pages} {1} (\bibinfo {year} {2020})}\BibitemShut
  {NoStop}%
\bibitem [{\citenamefont {Shen}(2016)}]{shen_fundamentals_2016}%
  \BibitemOpen
  \bibfield  {author} {\bibinfo {author} {\bibfnamefont {Y.~R.}\ \bibnamefont
  {Shen}},\ }\href {https://doi.org/10.1017/CBO9781316162613} {\emph {\bibinfo
  {title} {Fundamentals of {Sum}-{Frequency} {Spectroscopy}}}},\ Cambridge
  {Molecular} {Science}\ (\bibinfo  {publisher} {Cambridge University Press},\
  \bibinfo {address} {Cambridge},\ \bibinfo {year} {2016})\BibitemShut
  {NoStop}%
\bibitem [{\citenamefont {Raschke}\ \emph {et~al.}(2002)\citenamefont
  {Raschke}, \citenamefont {Hayashi}, \citenamefont {Lin},\ and\ \citenamefont
  {Shen}}]{raschke_doubly-resonant_2002}%
  \BibitemOpen
  \bibfield  {author} {\bibinfo {author} {\bibfnamefont {M.}~\bibnamefont
  {Raschke}}, \bibinfo {author} {\bibfnamefont {M.}~\bibnamefont {Hayashi}},
  \bibinfo {author} {\bibfnamefont {S.}~\bibnamefont {Lin}},\ and\ \bibinfo
  {author} {\bibfnamefont {Y.}~\bibnamefont {Shen}},\ }\bibfield  {title}
  {\bibinfo {title} {Doubly-resonant sum-frequency generation spectroscopy for
  surface studies},\ }\href {https://doi.org/10.1016/S0009-2614(02)00560-2}
  {\bibfield  {journal} {\bibinfo  {journal} {Chemical Physics Letters}\
  }\textbf {\bibinfo {volume} {359}},\ \bibinfo {pages} {367} (\bibinfo {year}
  {2002})}\BibitemShut {NoStop}%
\bibitem [{\citenamefont {Roke}\ and\ \citenamefont
  {Gonella}(2012)}]{roke_nonlinear_2012}%
  \BibitemOpen
  \bibfield  {author} {\bibinfo {author} {\bibfnamefont {S.}~\bibnamefont
  {Roke}}\ and\ \bibinfo {author} {\bibfnamefont {G.}~\bibnamefont {Gonella}},\
  }\bibfield  {title} {\bibinfo {title} {Nonlinear {Light} {Scattering} and
  {Spectroscopy} of {Particles} and {Droplets} in {Liquids}},\ }\href
  {https://doi.org/10.1146/annurev-physchem-032511-143748} {\bibfield
  {journal} {\bibinfo  {journal} {Annual Review of Physical Chemistry}\
  }\textbf {\bibinfo {volume} {63}},\ \bibinfo {pages} {353} (\bibinfo {year}
  {2012})}\BibitemShut {NoStop}%
\bibitem [{\citenamefont {Roelli}\ \emph {et~al.}(2016)\citenamefont {Roelli},
  \citenamefont {Galland}, \citenamefont {Piro},\ and\ \citenamefont
  {Kippenberg}}]{roelli_molecular_2016}%
  \BibitemOpen
  \bibfield  {author} {\bibinfo {author} {\bibfnamefont {P.}~\bibnamefont
  {Roelli}}, \bibinfo {author} {\bibfnamefont {C.}~\bibnamefont {Galland}},
  \bibinfo {author} {\bibfnamefont {N.}~\bibnamefont {Piro}},\ and\ \bibinfo
  {author} {\bibfnamefont {T.~J.}\ \bibnamefont {Kippenberg}},\ }\bibfield
  {title} {\bibinfo {title} {Molecular cavity optomechanics as a theory of
  plasmon-enhanced {Raman} scattering},\ }\href
  {https://doi.org/10.1038/nnano.2015.264} {\bibfield  {journal} {\bibinfo
  {journal} {Nature Nanotechnology}\ }\textbf {\bibinfo {volume} {11}},\
  \bibinfo {pages} {164} (\bibinfo {year} {2016})}\BibitemShut {NoStop}%
\bibitem [{\citenamefont {Shalabney}\ \emph {et~al.}(2015)\citenamefont
  {Shalabney}, \citenamefont {George}, \citenamefont {Hutchison}, \citenamefont
  {Pupillo}, \citenamefont {Genet},\ and\ \citenamefont
  {Ebbesen}}]{shalabney_coherent_2015}%
  \BibitemOpen
  \bibfield  {author} {\bibinfo {author} {\bibfnamefont {A.}~\bibnamefont
  {Shalabney}}, \bibinfo {author} {\bibfnamefont {J.}~\bibnamefont {George}},
  \bibinfo {author} {\bibfnamefont {J.}~\bibnamefont {Hutchison}}, \bibinfo
  {author} {\bibfnamefont {G.}~\bibnamefont {Pupillo}}, \bibinfo {author}
  {\bibfnamefont {C.}~\bibnamefont {Genet}},\ and\ \bibinfo {author}
  {\bibfnamefont {T.~W.}\ \bibnamefont {Ebbesen}},\ }\bibfield  {title}
  {\bibinfo {title} {Coherent coupling of molecular resonators with a
  microcavity mode},\ }\href {https://doi.org/10.1038/ncomms6981} {\bibfield
  {journal} {\bibinfo  {journal} {Nature Communications}\ }\textbf {\bibinfo
  {volume} {6}},\ \bibinfo {pages} {5981} (\bibinfo {year} {2015})}\BibitemShut
  {NoStop}%
\bibitem [{\citenamefont {Garcia-Vidal}\ \emph {et~al.}(2021)\citenamefont
  {Garcia-Vidal}, \citenamefont {Ciuti},\ and\ \citenamefont
  {Ebbesen}}]{garcia-vidal_manipulating_2021}%
  \BibitemOpen
  \bibfield  {author} {\bibinfo {author} {\bibfnamefont {F.~J.}\ \bibnamefont
  {Garcia-Vidal}}, \bibinfo {author} {\bibfnamefont {C.}~\bibnamefont
  {Ciuti}},\ and\ \bibinfo {author} {\bibfnamefont {T.~W.}\ \bibnamefont
  {Ebbesen}},\ }\bibfield  {title} {\bibinfo {title} {Manipulating matter by
  strong coupling to vacuum fields},\ }\bibfield  {journal} {\bibinfo
  {journal} {Science}\ }\textbf {\bibinfo {volume} {373}},\ \href
  {https://doi.org/10.1126/science.abd0336} {10.1126/science.abd0336} (\bibinfo
  {year} {2021})\BibitemShut {NoStop}%
\bibitem [{\citenamefont {Bylinkin}\ \emph {et~al.}(2021)\citenamefont
  {Bylinkin}, \citenamefont {Schnell}, \citenamefont {Autore}, \citenamefont
  {Calavalle}, \citenamefont {Li}, \citenamefont {Taboada-Gutièrrez},
  \citenamefont {Liu}, \citenamefont {Edgar}, \citenamefont {Casanova},
  \citenamefont {Hueso}, \citenamefont {Alonso-Gonzalez}, \citenamefont
  {Nikitin},\ and\ \citenamefont {Hillenbrand}}]{bylinkin_real-space_2021}%
  \BibitemOpen
  \bibfield  {author} {\bibinfo {author} {\bibfnamefont {A.}~\bibnamefont
  {Bylinkin}}, \bibinfo {author} {\bibfnamefont {M.}~\bibnamefont {Schnell}},
  \bibinfo {author} {\bibfnamefont {M.}~\bibnamefont {Autore}}, \bibinfo
  {author} {\bibfnamefont {F.}~\bibnamefont {Calavalle}}, \bibinfo {author}
  {\bibfnamefont {P.}~\bibnamefont {Li}}, \bibinfo {author} {\bibfnamefont
  {J.}~\bibnamefont {Taboada-Gutièrrez}}, \bibinfo {author} {\bibfnamefont
  {S.}~\bibnamefont {Liu}}, \bibinfo {author} {\bibfnamefont {J.~H.}\
  \bibnamefont {Edgar}}, \bibinfo {author} {\bibfnamefont {F.}~\bibnamefont
  {Casanova}}, \bibinfo {author} {\bibfnamefont {L.~E.}\ \bibnamefont {Hueso}},
  \bibinfo {author} {\bibfnamefont {P.}~\bibnamefont {Alonso-Gonzalez}},
  \bibinfo {author} {\bibfnamefont {A.~Y.}\ \bibnamefont {Nikitin}},\ and\
  \bibinfo {author} {\bibfnamefont {R.}~\bibnamefont {Hillenbrand}},\
  }\bibfield  {title} {\bibinfo {title} {Real-space observation of vibrational
  strong coupling between propagating phonon polaritons and organic
  molecules},\ }\href {https://doi.org/10.1038/s41566-020-00725-3} {\bibfield
  {journal} {\bibinfo  {journal} {Nature Photonics}\ }\textbf {\bibinfo
  {volume} {15}},\ \bibinfo {pages} {197} (\bibinfo {year} {2021})}\BibitemShut
  {NoStop}%
\bibitem [{\citenamefont {Liu}\ and\ \citenamefont
  {Shen}(2008)}]{liu_sum-frequency_2008}%
  \BibitemOpen
  \bibfield  {author} {\bibinfo {author} {\bibfnamefont {W.-T.}\ \bibnamefont
  {Liu}}\ and\ \bibinfo {author} {\bibfnamefont {Y.~R.}\ \bibnamefont {Shen}},\
  }\bibfield  {title} {\bibinfo {title} {Sum-frequency phonon spectroscopy on
  alpha-quartz},\ }\href {https://doi.org/10.1103/PhysRevB.78.024302}
  {\bibfield  {journal} {\bibinfo  {journal} {Physical Review B}\ }\textbf
  {\bibinfo {volume} {78}},\ \bibinfo {pages} {024302} (\bibinfo {year}
  {2008})}\BibitemShut {NoStop}%
\bibitem [{\citenamefont {Cremer}\ \emph {et~al.}(1996)\citenamefont {Cremer},
  \citenamefont {Su}, \citenamefont {Shen},\ and\ \citenamefont
  {Somorjai}}]{cremer_ethylene_1996}%
  \BibitemOpen
  \bibfield  {author} {\bibinfo {author} {\bibfnamefont {P.~S.}\ \bibnamefont
  {Cremer}}, \bibinfo {author} {\bibfnamefont {X.}~\bibnamefont {Su}}, \bibinfo
  {author} {\bibfnamefont {Y.~R.}\ \bibnamefont {Shen}},\ and\ \bibinfo
  {author} {\bibfnamefont {G.~A.}\ \bibnamefont {Somorjai}},\ }\bibfield
  {title} {\bibinfo {title} {Ethylene {Hydrogenation} on {Pt}(111) {Monitored}
  in {Situ} at {High} {Pressures} {Using} {Sum} {Frequency} {Generation}},\
  }\href {https://doi.org/10.1021/ja952800t} {\bibfield  {journal} {\bibinfo
  {journal} {Journal of the American Chemical Society}\ }\textbf {\bibinfo
  {volume} {118}},\ \bibinfo {pages} {2942} (\bibinfo {year}
  {1996})}\BibitemShut {NoStop}%
\bibitem [{\citenamefont {Hunt}\ \emph {et~al.}(1987)\citenamefont {Hunt},
  \citenamefont {Guyot-Sionnest},\ and\ \citenamefont
  {Shen}}]{hunt_observation_1987}%
  \BibitemOpen
  \bibfield  {author} {\bibinfo {author} {\bibfnamefont {J.~H.}\ \bibnamefont
  {Hunt}}, \bibinfo {author} {\bibfnamefont {P.}~\bibnamefont
  {Guyot-Sionnest}},\ and\ \bibinfo {author} {\bibfnamefont {Y.~R.}\
  \bibnamefont {Shen}},\ }\bibfield  {title} {\bibinfo {title} {Observation of
  {C}-{H} stretch vibrations of monolayers of molecules optical sum-frequency
  generation},\ }\href {https://doi.org/10.1016/0009-2614(87)87049-5}
  {\bibfield  {journal} {\bibinfo  {journal} {Chemical Physics Letters}\
  }\textbf {\bibinfo {volume} {133}},\ \bibinfo {pages} {189} (\bibinfo {year}
  {1987})}\BibitemShut {NoStop}%
\bibitem [{\citenamefont {Roke}\ \emph {et~al.}(2003)\citenamefont {Roke},
  \citenamefont {Schins}, \citenamefont {Müller},\ and\ \citenamefont
  {Bonn}}]{roke_vibrational_2003}%
  \BibitemOpen
  \bibfield  {author} {\bibinfo {author} {\bibfnamefont {S.}~\bibnamefont
  {Roke}}, \bibinfo {author} {\bibfnamefont {J.}~\bibnamefont {Schins}},
  \bibinfo {author} {\bibfnamefont {M.}~\bibnamefont {Müller}},\ and\ \bibinfo
  {author} {\bibfnamefont {M.}~\bibnamefont {Bonn}},\ }\bibfield  {title}
  {\bibinfo {title} {Vibrational {Spectroscopic} {Investigation} of the {Phase}
  {Diagram} of a {Biomimetic} {Lipid} {Monolayer}},\ }\href
  {https://doi.org/10.1103/PhysRevLett.90.128101} {\bibfield  {journal}
  {\bibinfo  {journal} {Physical Review Letters}\ }\textbf {\bibinfo {volume}
  {90}},\ \bibinfo {pages} {128101} (\bibinfo {year} {2003})}\BibitemShut
  {NoStop}%
\bibitem [{\citenamefont {Schnell}\ \emph {et~al.}(2010)\citenamefont
  {Schnell}, \citenamefont {Garcia-Etxarri}, \citenamefont {Alkorta},
  \citenamefont {Aizpurua},\ and\ \citenamefont
  {Hillenbrand}}]{schnell_phase-resolved_2010}%
  \BibitemOpen
  \bibfield  {author} {\bibinfo {author} {\bibfnamefont {M.}~\bibnamefont
  {Schnell}}, \bibinfo {author} {\bibfnamefont {A.}~\bibnamefont
  {Garcia-Etxarri}}, \bibinfo {author} {\bibfnamefont {J.}~\bibnamefont
  {Alkorta}}, \bibinfo {author} {\bibfnamefont {J.}~\bibnamefont {Aizpurua}},\
  and\ \bibinfo {author} {\bibfnamefont {R.}~\bibnamefont {Hillenbrand}},\
  }\bibfield  {title} {\bibinfo {title} {Phase-{Resolved} {Mapping} of the
  {Near}-{Field} {Vector} and {Polarization} {State} in {Nanoscale} {Antenna}
  {Gaps}},\ }\href {https://doi.org/10.1021/nl101693a} {\bibfield  {journal}
  {\bibinfo  {journal} {Nano Letters}\ }\textbf {\bibinfo {volume} {10}},\
  \bibinfo {pages} {3524} (\bibinfo {year} {2010})}\BibitemShut {NoStop}%
\bibitem [{\citenamefont {Ahmed}\ \emph {et~al.}(2021)\citenamefont {Ahmed},
  \citenamefont {Banjac}, \citenamefont {Verlekar}, \citenamefont {Cometto},
  \citenamefont {Lingenfelder},\ and\ \citenamefont
  {Galland}}]{ahmed_structural_2021}%
  \BibitemOpen
  \bibfield  {author} {\bibinfo {author} {\bibfnamefont {A.}~\bibnamefont
  {Ahmed}}, \bibinfo {author} {\bibfnamefont {K.}~\bibnamefont {Banjac}},
  \bibinfo {author} {\bibfnamefont {S.~S.}\ \bibnamefont {Verlekar}}, \bibinfo
  {author} {\bibfnamefont {F.~P.}\ \bibnamefont {Cometto}}, \bibinfo {author}
  {\bibfnamefont {M.}~\bibnamefont {Lingenfelder}},\ and\ \bibinfo {author}
  {\bibfnamefont {C.}~\bibnamefont {Galland}},\ }\bibfield  {title} {\bibinfo
  {title} {Structural {Order} of the {Molecular} {Adlayer} {Impacts} the
  {Stability} of {Nanoparticle}-on-{Mirror} {Plasmonic} {Cavities}},\
  }\bibfield  {journal} {\bibinfo  {journal} {ACS Photonics}\ }\href
  {https://doi.org/10.1021/acsphotonics.1c00645} {10.1021/acsphotonics.1c00645}
  (\bibinfo {year} {2021})\BibitemShut {NoStop}%
\bibitem [{\citenamefont {Zhang}\ \emph {et~al.}(2013)\citenamefont {Zhang},
  \citenamefont {Zhang}, \citenamefont {Dong}, \citenamefont {Jiang},
  \citenamefont {Zhang}, \citenamefont {Chen}, \citenamefont {Zhang},
  \citenamefont {Liao}, \citenamefont {Aizpurua}, \citenamefont {Luo},
  \citenamefont {Yang},\ and\ \citenamefont {Hou}}]{zhang_chemical_2013}%
  \BibitemOpen
  \bibfield  {author} {\bibinfo {author} {\bibfnamefont {R.}~\bibnamefont
  {Zhang}}, \bibinfo {author} {\bibfnamefont {Y.}~\bibnamefont {Zhang}},
  \bibinfo {author} {\bibfnamefont {Z.~C.}\ \bibnamefont {Dong}}, \bibinfo
  {author} {\bibfnamefont {S.}~\bibnamefont {Jiang}}, \bibinfo {author}
  {\bibfnamefont {C.}~\bibnamefont {Zhang}}, \bibinfo {author} {\bibfnamefont
  {L.~G.}\ \bibnamefont {Chen}}, \bibinfo {author} {\bibfnamefont
  {L.}~\bibnamefont {Zhang}}, \bibinfo {author} {\bibfnamefont
  {Y.}~\bibnamefont {Liao}}, \bibinfo {author} {\bibfnamefont {J.}~\bibnamefont
  {Aizpurua}}, \bibinfo {author} {\bibfnamefont {Y.}~\bibnamefont {Luo}},
  \bibinfo {author} {\bibfnamefont {J.~L.}\ \bibnamefont {Yang}},\ and\
  \bibinfo {author} {\bibfnamefont {J.~G.}\ \bibnamefont {Hou}},\ }\bibfield
  {title} {\bibinfo {title} {Chemical mapping of a single molecule by
  plasmon-enhanced {Raman} scattering},\ }\href
  {https://doi.org/10.1038/nature12151} {\bibfield  {journal} {\bibinfo
  {journal} {Nature}\ }\textbf {\bibinfo {volume} {498}},\ \bibinfo {pages}
  {82} (\bibinfo {year} {2013})}\BibitemShut {NoStop}%
\bibitem [{\citenamefont {Zhang}\ \emph {et~al.}(1994)\citenamefont {Zhang},
  \citenamefont {Gutow},\ and\ \citenamefont
  {Eisenthal}}]{zhang_vibrational_1994}%
  \BibitemOpen
  \bibfield  {author} {\bibinfo {author} {\bibfnamefont {D.}~\bibnamefont
  {Zhang}}, \bibinfo {author} {\bibfnamefont {J.}~\bibnamefont {Gutow}},\ and\
  \bibinfo {author} {\bibfnamefont {K.~B.}\ \bibnamefont {Eisenthal}},\
  }\bibfield  {title} {\bibinfo {title} {Vibrational {Spectra}, {Orientations},
  and {Phase} {Transitions} in {Long}-{Chain} {Amphiphiles} at the
  {Air}/{Water} {Interface}: {Probing} the {Head} and {Tail} {Groups} by {Sum}
  {Frequency} {Generation}},\ }\href {https://doi.org/10.1021/j100102a045}
  {\bibfield  {journal} {\bibinfo  {journal} {The Journal of Physical
  Chemistry}\ }\textbf {\bibinfo {volume} {98}},\ \bibinfo {pages} {13729}
  (\bibinfo {year} {1994})}\BibitemShut {NoStop}%
\bibitem [{\citenamefont {Wang}\ \emph {et~al.}(2013)\citenamefont {Wang},
  \citenamefont {Yan},\ and\ \citenamefont {Chen}}]{wang_sers_2013}%
  \BibitemOpen
  \bibfield  {author} {\bibinfo {author} {\bibfnamefont {Y.}~\bibnamefont
  {Wang}}, \bibinfo {author} {\bibfnamefont {B.}~\bibnamefont {Yan}},\ and\
  \bibinfo {author} {\bibfnamefont {L.}~\bibnamefont {Chen}},\ }\bibfield
  {title} {\bibinfo {title} {{SERS} {Tags}: {Novel} {Optical} {Nanoprobes} for
  {Bioanalysis}},\ }\href {https://doi.org/10.1021/cr300120g} {\bibfield
  {journal} {\bibinfo  {journal} {Chemical Reviews}\ }\textbf {\bibinfo
  {volume} {113}},\ \bibinfo {pages} {1391} (\bibinfo {year}
  {2013})}\BibitemShut {NoStop}%
\bibitem [{\citenamefont {Rodrigo}\ \emph {et~al.}(2015)\citenamefont
  {Rodrigo}, \citenamefont {Limaj}, \citenamefont {Janner}, \citenamefont
  {Etezadi}, \citenamefont {Abajo}, \citenamefont {Pruneri},\ and\
  \citenamefont {Altug}}]{rodrigo_mid-infrared_2015}%
  \BibitemOpen
  \bibfield  {author} {\bibinfo {author} {\bibfnamefont {D.}~\bibnamefont
  {Rodrigo}}, \bibinfo {author} {\bibfnamefont {O.}~\bibnamefont {Limaj}},
  \bibinfo {author} {\bibfnamefont {D.}~\bibnamefont {Janner}}, \bibinfo
  {author} {\bibfnamefont {D.}~\bibnamefont {Etezadi}}, \bibinfo {author}
  {\bibfnamefont {F.~J. G.~d.}\ \bibnamefont {Abajo}}, \bibinfo {author}
  {\bibfnamefont {V.}~\bibnamefont {Pruneri}},\ and\ \bibinfo {author}
  {\bibfnamefont {H.}~\bibnamefont {Altug}},\ }\bibfield  {title} {\bibinfo
  {title} {Mid-infrared plasmonic biosensing with graphene},\ }\href
  {https://doi.org/10.1126/science.aab2051} {\bibfield  {journal} {\bibinfo
  {journal} {Science}\ }\textbf {\bibinfo {volume} {349}},\ \bibinfo {pages}
  {165} (\bibinfo {year} {2015})}\BibitemShut {NoStop}%
\bibitem [{\citenamefont {Epstein}\ \emph {et~al.}(2020)\citenamefont
  {Epstein}, \citenamefont {Alcaraz}, \citenamefont {Huang}, \citenamefont
  {Pusapati}, \citenamefont {Hugonin}, \citenamefont {Kumar}, \citenamefont
  {Deputy}, \citenamefont {Khodkov}, \citenamefont {Rappoport}, \citenamefont
  {Hong}, \citenamefont {Peres}, \citenamefont {Kong}, \citenamefont {Smith},\
  and\ \citenamefont {Koppens}}]{epstein_far-field_2020}%
  \BibitemOpen
  \bibfield  {author} {\bibinfo {author} {\bibfnamefont {I.}~\bibnamefont
  {Epstein}}, \bibinfo {author} {\bibfnamefont {D.}~\bibnamefont {Alcaraz}},
  \bibinfo {author} {\bibfnamefont {Z.}~\bibnamefont {Huang}}, \bibinfo
  {author} {\bibfnamefont {V.-V.}\ \bibnamefont {Pusapati}}, \bibinfo {author}
  {\bibfnamefont {J.-P.}\ \bibnamefont {Hugonin}}, \bibinfo {author}
  {\bibfnamefont {A.}~\bibnamefont {Kumar}}, \bibinfo {author} {\bibfnamefont
  {X.~M.}\ \bibnamefont {Deputy}}, \bibinfo {author} {\bibfnamefont
  {T.}~\bibnamefont {Khodkov}}, \bibinfo {author} {\bibfnamefont {T.~G.}\
  \bibnamefont {Rappoport}}, \bibinfo {author} {\bibfnamefont {J.-Y.}\
  \bibnamefont {Hong}}, \bibinfo {author} {\bibfnamefont {N.~M.~R.}\
  \bibnamefont {Peres}}, \bibinfo {author} {\bibfnamefont {J.}~\bibnamefont
  {Kong}}, \bibinfo {author} {\bibfnamefont {D.~R.}\ \bibnamefont {Smith}},\
  and\ \bibinfo {author} {\bibfnamefont {F.~H.~L.}\ \bibnamefont {Koppens}},\
  }\bibfield  {title} {\bibinfo {title} {Far-field excitation of single
  graphene plasmon cavities with ultracompressed mode volumes},\ }\href
  {https://doi.org/10.1126/science.abb1570} {\bibfield  {journal} {\bibinfo
  {journal} {Science}\ }\textbf {\bibinfo {volume} {368}},\ \bibinfo {pages}
  {1219} (\bibinfo {year} {2020})}\BibitemShut {NoStop}%
\bibitem [{\citenamefont {Tittl}\ \emph {et~al.}(2018)\citenamefont {Tittl},
  \citenamefont {Leitis}, \citenamefont {Liu}, \citenamefont {Yesilkoy},
  \citenamefont {Choi}, \citenamefont {Neshev}, \citenamefont {Kivshar},\ and\
  \citenamefont {Altug}}]{tittl_imaging-based_2018}%
  \BibitemOpen
  \bibfield  {author} {\bibinfo {author} {\bibfnamefont {A.}~\bibnamefont
  {Tittl}}, \bibinfo {author} {\bibfnamefont {A.}~\bibnamefont {Leitis}},
  \bibinfo {author} {\bibfnamefont {M.}~\bibnamefont {Liu}}, \bibinfo {author}
  {\bibfnamefont {F.}~\bibnamefont {Yesilkoy}}, \bibinfo {author}
  {\bibfnamefont {D.-Y.}\ \bibnamefont {Choi}}, \bibinfo {author}
  {\bibfnamefont {D.~N.}\ \bibnamefont {Neshev}}, \bibinfo {author}
  {\bibfnamefont {Y.~S.}\ \bibnamefont {Kivshar}},\ and\ \bibinfo {author}
  {\bibfnamefont {H.}~\bibnamefont {Altug}},\ }\bibfield  {title} {\bibinfo
  {title} {Imaging-based molecular barcoding with pixelated dielectric
  metasurfaces},\ }\href {https://doi.org/10.1126/science.aas9768} {\bibfield
  {journal} {\bibinfo  {journal} {Science}\ }\textbf {\bibinfo {volume}
  {360}},\ \bibinfo {pages} {1105} (\bibinfo {year} {2018})}\BibitemShut
  {NoStop}%
\bibitem [{\citenamefont {Mott}\ and\ \citenamefont
  {Rez}(2012)}]{mott_calculated_2012}%
  \BibitemOpen
  \bibfield  {author} {\bibinfo {author} {\bibfnamefont {A.~J.}\ \bibnamefont
  {Mott}}\ and\ \bibinfo {author} {\bibfnamefont {P.}~\bibnamefont {Rez}},\
  }\bibfield  {title} {\bibinfo {title} {Calculated infrared spectra of nerve
  agents and simulants},\ }\href {https://doi.org/10.1016/j.saa.2012.02.010}
  {\bibfield  {journal} {\bibinfo  {journal} {Spectrochimica Acta Part A:
  Molecular and Biomolecular Spectroscopy}\ }\textbf {\bibinfo {volume} {91}},\
  \bibinfo {pages} {256} (\bibinfo {year} {2012})}\BibitemShut {NoStop}%
\end{thebibliography}%

\end{document}